\documentclass[preprint]{aastex}
\usepackage{natbib}
\usepackage{emulateapj5}
\submitted{To appear in the Astrophysical Journal, Letters}
\shorttitle{DIFFUSE FUV EMISSION}
\shortauthors{DIXON ET AL.}

\begin{document}
 

\newcommand{\cf}{cf.}
\newcommand{\eg}{e.g.}
\newcommand{\etal}{et~al.}
\newcommand{\etc}{etc.}
\newcommand{\fig}[1]{Fig.~\ref{#1}}
\newcommand{\h}{$^{\rm h}$}
\newcommand{\ie}{i.e.}
\newcommand{\m}{$^{\rm m}$}

\newcommand{\cfour}{\ion{C}{4}}
\newcommand{\ebv}{$E($\bv)}
\newcommand{\f}{{\it f}\/}
\newcommand{\hone}{\ion{H}{1}}
\newcommand{\htwo}{H$_2$}
\newcommand{\kms}{km s$^{-1}$}
\newcommand{\lu}{photons cm$^{-2}$ s$^{-1}$ sr$^{-1}$}
\newcommand{\nho}{$N$(\hone)}
\newcommand{\nht}{$N$(H$_2$)}
\newcommand{\nosix}{$N$(\osix)}
\newcommand{\oone}{\ion{O}{1}}
\newcommand{\osix}{\ion{O}{6}}
\newcommand{\rv}{$R_V$}
\newcommand{\specfit}{{\small SPECFIT}}

\newcommand{\chips}{{\it CHIPS}}
\newcommand{\cop}{{\it Copernicus}}
\newcommand{\euve}{{\it EUVE}}
\newcommand{\fuse}{{\it FUSE}}
\newcommand{\hst}{{\it HST}}
\newcommand{\hut}{HUT}
\newcommand{\iue}{{\it IUE}}
\newcommand{\orf}{{ORFEUS}}
\newcommand{\rosat}{{\it ROSAT}}

\newcommand{\na}{New~Astronomy}

\title{FUSE Detection of Diffuse Galactic O VI Emission toward the Coma
and Virgo Clusters\footnotemark}

\footnotetext[1]{Based on observations made with the NASA-CNES-CSA {\it
Far Ultraviolet Spectroscopic Explorer.  FUSE} is operated for NASA by
the Johns Hopkins University under NASA contract NAS5-32985.}

\author{W.\ Van Dyke Dixon\altaffilmark{2}, Shauna Sallmen, and Mark Hurwitz}
\affil{Space Sciences Laboratory, University of California, Berkeley, Berkeley, CA 94720-7450;}

\email{vand@ssl.berkeley.edu, sallmen@ssl.berkeley.edu, markh@ssl.berkeley.edu}

\altaffiltext{2}{Current address: Department of Physics and Astronomy,
The Johns Hopkins University, Baltimore, MD 21218}

\and

\author{Richard Lieu}
\affil{Department of Physics, University of Alabama in Huntsville, Huntsville, AL 35899;
lieur@cspar.uah.edu}


\begin{abstract}

We report the detection of diffuse \osix\ $\lambda \lambda 1032, 1038$
emission in a 29-ksec observation centered on the Coma Cluster ($l =
57.6, b = +88.0$) and an 11-ksec observation toward Virgo ($l = 284.2,
b = +74.5$) through the 30\arcsec\ $\times$ 30\arcsec\ aperture
of the {\it Far Ultraviolet Spectroscopic Explorer (FUSE).} The
emission lines have a redshift near zero and are thus produced by gas
in our own Galaxy.  Observed surface brightnesses are $2000 \pm 600$
photons cm$^{-2}$ s$^{-1}$ sr$^{-1}$ for each of the \osix\ components
in the Coma spectrum, and $2900 \pm 700$ and $1700 \pm 700$ photons
cm$^{-2}$ s$^{-1}$ sr$^{-1}$ for the 1032 and 1038 \AA\ lines,
respectively, toward Virgo.  These features are similar in strength to
those recently observed in the southern Galactic hemisphere ($l =
315.0, b =-41.3$) in an $\sim$ 200 ksec \fuse\ observation.  From a
\fuse\ spectrum of M87, we find that $N($\osix) toward Virgo is $(1.4
\pm 0.8) \times 10^{14}$ cm$^{-2}$.  By combining emission- and
absorption-line data for this sight line, we estimate the physical
parameters of the emitting gas.

\end{abstract}
 

\section{INTRODUCTION}

Emission spectroscopy at extreme- and far-ultraviolet wavelengths will
provide the key to our eventual understanding of the origin,
distribution, and physical processes of the hot interstellar medium
(ISM).  Observing diffuse interstellar emission at these wavelengths
has, however, proven technically challenging, and only a handful of
detections have been obtained to date \citep*{Martin:Bowyer:90,
Dixon:96}.  Thus, the recent observation of diffuse Galactic
\osix\ emission with the {\it Far Ultraviolet Spectroscopic Explorer
(FUSE)} by \citet{Shelton:01} is both exciting,
because of what it can tell us about the state and distribution of hot
gas in the Galaxy, and frustrating, because of the long integration
times ($\sim 200$ ksec) required.  We are therefore pleased to report
the serendipitous detection of diffuse Galactic \osix\ emission in two
(relatively) short \fuse\ observations of the Coma and Virgo Clusters.
Combining our emission-line intensities toward Virgo with the
\osix\ column density measured toward M87, we constrain the physical
properties of the emitting gas.  These results bode well for future
efforts to map the sky at far-UV (FUV) wavelengths.

\section{OBSERVATIONS AND DATA REDUCTION}

\fuse\ comprises four separate optical systems.  Two employ LiF optical
coatings and are sensitive to wavelengths from 990 to 1187 \AA, while
the other two use SiC coatings, which provide reflectivity to
wavelengths as short as 905 \AA.  The four channels overlap between 990
and 1070 \AA.  For a complete description of \fuse, see \citet{Moos:00}
and \citet{Sahnow:00}.

The \fuse\ spectrum of the Coma Cluster was obtained in 17 separate
exposures on 2000 June 18 and 19.  Each exposure was centered on 12\h
59\m 49\fs 0, +27\arcdeg 57\arcmin 46\arcsec\ (J2000, or $l = 57.61, b
= +87.96$ in Galactic coordinates), near the center of the Coma
Cluster.  The total exposure time was 28608 s, with 23553 s obtained
during orbital night.  We use the entire 29-ksec data set in our
analysis.  Our spectrum of the Virgo Cluster is a combination of data
from two locations near the cluster center.  Two exposures, centered on
12\h 31\m 07\fs 3, +12\arcdeg 23\arcmin 46\arcsec\ ($l = 284.03, b =
+74.52$) and totaling 2242 s, were obtained on 2000 June 13.  Seven
exposures, centered on 12\h 31\m 13\fs 4, +12\arcdeg 22\arcmin
10\arcsec\ ($l = 284.15, b = +74.50$) and totaling 8688 s, were
obtained on 2000 June 17.  The total integration time is 10,930 s, all
of it during orbital night.  All observations were made through the
30\arcsec\ $\times$ 30\arcsec\ low-resolution (LWRS) aperture.  The
data-reduction procedures applied to these data are discussed in detail
in \citet{ClusterOVI} and are not repeated here.

While the resolution of the \fuse\ spectrograph is approximately 15
\kms\ for a point source,  diffuse emission filling the LWRS aperture
yields a line profile that is well approximated by a top-hat function
with a width of $\sim$ 106 \kms.  Our spectra are quite faint, so we
bin the Coma data by 8 detector pixels, or about 15.5 \kms.
Approximately 7 of these bins are required to span a diffuse emission
feature, providing ample sensitivity to its line profile.  We bin the
data from the shorter Virgo observation by 16 pixels.  We distinguish
between detector pixels and 8- or 16-pixel bins throughout this
Letter.  Because of the wide disparity in the effective areas of the
various detector
\linebreak[4]

\includegraphics{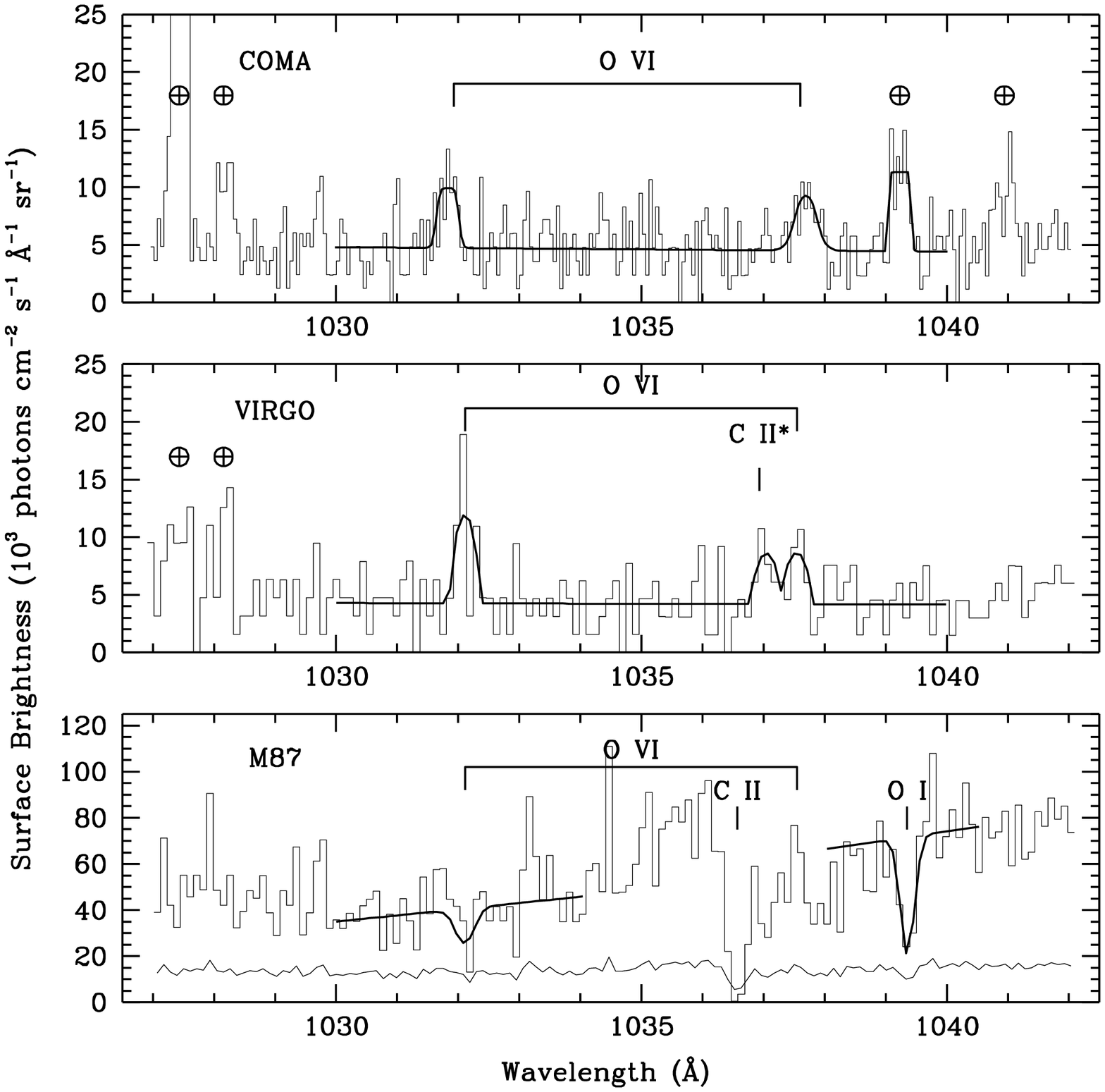}
\vspace*{3.3in}
\figcaption{\label{spectra}
\fuse\ spectra of the {\it (top)} Coma and {\it (middle)}
Virgo Clusters showing
diffuse Galactic O~VI $\lambda \lambda 1032, 1038$ emission.  The Virgo
spectrum also shows Galactic C~II* $\lambda 1037.02$ emission.
Terrestrial airglow features are marked with $\earth$.  The Coma
spectrum is binned by 8 detector pixels, the Virgo spectrum by 16.  The
data, from detector LiF 1A,  are presented as histograms and are
overplotted by our best-fit models.  The observed continuum level is
consistent with the dark-count rate determined from unilluminated
regions of the detector.  The round shape of the model emission
features in the Virgo spectrum (which are in fact top-hat functions) is
an artifact of our sparse sampling of the model lines.  Bottom:
\fuse\ spectrum of M87, a weighted mean of the LiF~1A and LiF~2B
spectra, binned by 16 detector pixels.  The error spectrum is
overplotted.  Heavy lines are the best-fit O~VI absorption model,
with $N$(O~VI) = $1.4 \times 10^{14}$ cm$^{-2}$, and a Gaussian fit to
the O~I $\lambda 1039.23$ line, which we adopt as the absorption
profile for O VI.}
\vskip 0.3in

\noindent
segments \citep{Sahnow:00}, combining faint spectra from different
segments does not significantly improve their signal-to-noise ratio.
We thus use data only from the segment with the highest effective area
at the wavelength of interest.

The \fuse\ flux calibration, based on theoretical models of white-dwarf
stellar atmospheres, is believed accurate to about 10\%
\citep{Sahnow:00}.  Corrections to the nominal \fuse\ wavelength scale
are derived from the measured positions of airglow features in the
day-time Coma spectrum and are good to about 0.01 \AA.  Error bars are
assigned to the data assuming Gaussian statistics, then smoothed by 9
bins to remove small-scale features in the error spectrum without
significantly changing its shape.  Segments of the flux- and
wavelength-calibrated LiF~1A spectra of each cluster, showing the
region about \osix\ $\lambda \lambda 1032, 1038$, are presented in
\fig{spectra}.  Upper limits on the redshifted emission from both
clusters are presented in \citet{ClusterOVI}.

\section{SPECTRAL ANALYSIS}

\footnotetext[3]{The Image Reduction and Analysis Facility (IRAF) is
distributed by the National Optical Astronomy Observatories, which is
operated by the Association of Universities for Research in Astronomy,
Inc., (AURA) under cooperative agreement with the National Science
Foundation.}

In our analysis, model spectra are fit to the flux-calibrated data
using the nonlinear curve-fitting program
\linebreak[4]

\vskip -25pt
\vbox to 2.5in {
\begin{center}
\small
{\sc Table 1\\
\osix\ Emission Features}
\vskip 2pt
\footnotesize
\begin{tabular}{cccc}
\tableline
\tableline
& Surface & Intrinsic & LSR \\
Wavelength & Brightness & FWHM & Velocity \\
(\AA) & (ph cm$^{-2}$ s$^{-1}$ sr$^{-1}$) & (\kms) & (\kms) \\
\tableline
\multicolumn{4}{c}{Coma} \\
\tableline
$1031.83 \pm 0.03$ & $2000 \pm 600$ & $23 \pm 55$ & $-17 \pm \phn9$ \\
$1037.70 \pm 0.06$ & $2000 \pm 600$ & $75 \pm 60$ & $+35 \pm 17$ \\
\tableline
\multicolumn{4}{c}{Virgo} \\
\tableline
$1032.11 \pm 0.05$ & $2900 \pm 700$ & $<80$ & $+84 \pm 15$ \\
$1037.55 \pm 0.06$ & $1700 \pm 700$ & $<110$ & $+12 \pm 17$ \\
\tableline
\end{tabular}
\vskip 2pt
\parbox{3.5in}{
\small\baselineskip 9pt
\footnotesize
\indent
{\sc Note.}--- Error bars are statistical and represent a change in
$\chi^2$ of 1.0.  Uncertainties in the area of the LWRS aperture and the
\fuse\ flux calibration contribute an additional 14\% systematic
uncertainty to the quoted line fluxes.  For lines narrower than our
top-hat function, 1-$\sigma$ upper limits to the intrinsic FWHM are quoted.}
\end{center}
}
 
\noindent
\specfit\ \citep{Kriss:94}, which runs in the IRAF\footnotemark\
environment, to perform a $\chi^2$
minimization of the model parameters.  The observed profile of a
diffuse emission feature represents a convolution of its intrinsic
profile, assumed to be a Gaussian, with the 106 \kms\ top-hat function
discussed above.  Free parameters in the fit are the level and slope of
the continuum (assumed linear) and the intensity, wavelength, and
intrinsic FWHM of each emission line.  The observed continuum level is
consistent with the dark-count rate determined from unilluminated
regions of the detector \citep{ClusterOVI}.

The model that best fits our Coma spectrum between 1030 and 1040 \AA\
yields a $\chi^2$ of 166.7 for 173 degrees of freedom, including the
\oone\ $\lambda 1039.23$ airglow line.  The Virgo spectrum, taken at
night, does not show the \oone\ feature, but does show emission from
\ion{C}{2}* $\lambda 1037.02$.  (Significant \ion{C}{2} $\lambda
1036.34$ emission is not expected; \citealt{Shelton:01}).  The best-fit
model returns a $\chi^2$ of 80.4 for 85 degrees of freedom.  These
models are overplotted on the data in \fig{spectra}.  Parameters of the
best-fit \osix\ lines are presented in Table 1, those of the
\ion{C}{2}* line in Table 2.  In both tables, error bars for each model
parameter are determined by increasing the best-fit value of that
parameter, while re-optimizing the other model parameters, until
$\chi^2$ increases by 1.0 (corresponding to a 1-$\sigma$ deviation for
one interesting parameter; \citealt{Avni:76}).  Because our wavelength
scale is based on the measured positions of airglow lines, derived
velocities are geocentric, and conversion to the Local Standard of Rest
(LSR; \citealt{Mihalas:Binney:81}) is straightforward.

We use the arguments of \citet{Shelton:01} to reject four alternative
explanations for the emission observed in the 1030--1040 \AA\ region of
the LiF 1A spectrum: scattered light in the cross-dispersion direction
on the detector, atmospheric airglow emission, solar
\osix\ contamination, and \htwo\ fluorescence.  Furthermore, none of
the observed features can be identified as redshifted emission from
either background cluster.  We conclude that the observed features
represent diffuse Galactic emission.

\begin{table*}
\begin{center}
\vskip 3pt
\small
{\sc Table 2\\
Additional Spectral Features}
\vskip 2pt
\footnotesize
\begin{tabular}{ccccc}
\tableline
\tableline
Wavelength & Surface Brightness & Intrinsic FWHM & Detector \\
(\AA) & (ph cm$^{-2}$ s$^{-1}$ sr$^{-1}$) & (\kms) & Segment & Identification \\
\tableline
\multicolumn{5}{c}{Coma} \\
\tableline
$\phn977.04 \pm 0.08$   & $3500 \pm 2400$       & $\phn43 \pm 140$ & SiC 2A & C III $\lambda 977.020$ \\
$1014.92 \pm 0.09$      & $2600 \pm \phn700$    & $120 \pm \phn50$ & LiF 1A & S III $\lambda 1015.505$ \\
$1122.17 \pm 0.04$      & $1600 \pm \phn500$    & $<40$         & LiF 2A & C I $\lambda 1122.260$ \\
$1138.96 \pm 0.02$      & $1800 \pm \phn500$    & $<15$         & LiF 2A & Artifact?\\
$1144.94 \pm 0.06$      & $2000 \pm \phn600$    & $<50$         & LiF 2A & Fe II $\lambda 1144.938$ \\
\tableline
\multicolumn{5}{c}{Virgo} \\
\tableline
$\phn975.17 \pm 0.04$   & $6400 \pm 1500$       & $<40$         & SiC 2A & \multicolumn{1}{c}{\nodata} \\
$\phn977.15 \pm 0.12$   & $7700 \pm 3300$       & $117 \pm 130$ & SiC 2A & C III $\lambda 977.020$ \\
$1037.04 \pm 0.07$      & $1700 \pm \phn700$    & $<110$        & LiF 1A & C II* $\lambda 1037.018$ \\
\tableline
\end{tabular}
\vskip 2pt
\parbox{4.7in}{
\small\baselineskip 9pt
\footnotesize
\indent
{\sc Note.}--- See note to Table 1.  To minimize contamination 
by solar emission, upper limits for the SiC~2A channel are
derived only from data obtained during orbital night.}
\end{center}
\vskip -10pt
\end{table*}

To search for statistically-significant emission at other wavelengths,
we bin each spectrum to the instrument resolution and identify
individual resolution elements with a signal-to-noise ratio greater
than 3.  Each such peak corresponds to a single emission feature.
Returning to our nominally-binned spectrum, we fit each feature with a
linear continuum and model emission line.  Because our data are rather
noisy, we employ two measures of a feature's statistical significance.
First, we determine the 1-$\sigma$ uncertainty in its measured flux
using the $\Delta \chi^2 = 1$ test described above; second, we test
whether deleting the feature from our best-fit model raises $\chi^2$ by
at least 9 \citep[corresponding to a 3-$\sigma$ deviation for one
interesting parameter, in this case the flux in the line;][]{Avni:76}.
We list in Table 2 those features that are significant
at the 3-$\sigma$ level by both tests, with three exceptions:  First,
we discard any features detected in known stray-light stripes (see
\citealt{ClusterOVI}).  Second, we include the \ion{C}{2}* $\lambda
1037$ line in the Virgo spectrum to facilitate comparison with a
similar feature seen by \citet{Shelton:01}.  Third, we include the
\ion{C}{3} $\lambda 977$ feature, not seen by \citeauthor{Shelton:01},
because it is present in both of our spectra.

Two lines presented in Table 2 remain unidentified.
Because it is so narrow, we suggest that the $\lambda 1138.96$ feature
in the Coma spectrum may represent a statistical fluctuation or detector
artifact.  The $\lambda 975.17$ feature in the Virgo spectrum might be
redshifted Ly $\gamma$, but its wavelength is too short: the line would
appear at 976.04 \AA\ at the redshift of Virgo ($z = 0.0036$;
\citealt{Ebeling:98}) and 976.78 \AA\ at that of M87 ($z = 0.00436$;
\citealt{Smith:00}).  Furthermore, most excitation mechanisms would
produce more Ly $\beta$ than Ly $\gamma$ emission, and redshifted Ly
$\beta$ is not seen in this spectrum \citep{ClusterOVI}.

\section{N(O \small{VI}) TOWARDS M87}

Given the emission intensity and column density of an emitting species
along a given line of sight, one can derive the electron density and
thermal pressure of the emitting plasma \citep{Shull:Slavin:94}.  We have
the \osix\ intensity toward Virgo; to constrain its column density, we
consider a \fuse\ spectrum of M87, the central galaxy of the Virgo
Cluster.

Two night-time exposures, totaling 3611 s, were obtained on 2000 June
13 with the LWRS aperture centered on 12\h 30\m 49\fs 1, +12\arcdeg
23\arcmin 31\arcsec\ ($l = 283.77, b = +74.49$), about 7 arcsec from
the galaxy's core.  To reduce the data, we employ the CALFUSE
pipeline \citep*{Sahnow:00, Oegerle:00}, version 1.8.7, making two
changes to the default parameters: first, we apply the pulse-height cut
used for our cluster data, excluding photon events with pulse heights
less than 4 or greater than 15 in standard arbitrary units; second,
we rescale the assumed background
count rate (which depends on the pulse-height cut) until the residual
continuum in the high- and medium-resolution apertures (which are
essentially blank fields) is no more than 1--2\% of the continuum in the LWRS
aperture.  To maximize the signal-to-noise ratio, the LiF~1A and LiF~2B
spectra are shifted to a common wavelength scale (using the correction
derived for Coma; see \citealt{ClusterOVI}),
weighted by their respective effective areas,
averaged, and binned by 16 detector pixels.  The resulting spectrum is
presented in \fig{spectra}.  The strongest absorption features
are due to Galactic \ion{C}{2} $\lambda 1036.34$ and \ion{O}{1}
$\lambda 1039.23$.

We generate a set of synthetic \osix\ absorption-line models using an
ISM line-fitting package written at U.C. Berkeley by M.\ Hurwitz and
V.\ Saba.  Wavelengths, oscillator strengths, and other atomic data are
taken from \citet{Morton:91}.  Given the column density, Doppler
broadening parameter, and a line list for each component, the program
computes a Voigt profile for each absorption feature and outputs a
high-resolution spectrum of $\tau$ versus wavelength.  We adopt the
Doppler parameter $b = 23$ \kms\ appropriate for a $5 \times 10^5$ K
gas.  The model spectra are convolved with a Gaussian of FWHM = 80
\kms\ (the observed width of the \ion{O}{1} $\lambda 1039.23$ feature)
and rebinned to the resolution of our M87 spectrum.

Using \specfit, we fit our synthetic absorption profiles and a linear
continuum to the 1030--1034 \AA\ region of our M87 spectrum.  The
wavelength of the absorption feature is fixed at 1032.11 \AA, the
observed wavelength of the \osix\ $\lambda 1032$ emission in the Virgo
spectrum.  The best-fit model (plotted in \fig{spectra}) corresponds to
a column density \nosix\ = $(1.4 \pm 0.8) \times 10^{14}$ cm$^{-2}$,
where the error bars represent $\Delta \chi^2 = 1$.  The feature is
statistically significant at the 2-$\sigma$ level, but appears both
weaker and narrower than the \ion{C}{2} and \ion{O}{1} lines.  Our
quoted column density may best be considered an upper limit.

Correcting the observed \osix\ doublet intensity $I_o \sim 5000$
\lu\ for scattering due to dust (a factor of about 1.5; see below) and
self-absorption within the emitting cloud (a factor of about 2, though
$\tau$ differs for the two components; see \citealt{Shelton:01}),
yields an intrinsic \osix\ intensity toward Virgo of $I_i \sim 15,000$
\lu.  Using equation (5) of \citet{Shull:Slavin:94}, we combine $I_i$
with \nosix\ towards M87 to derive the electron density $n_e$ of the
emitting plasma.  Over the temperature range for which \osix\ is an
important plasma constituent, $5.3 < \log T < 5.8$, $n_e$ is nearly
constant, varying between 0.046 and 0.048 cm$^{-3}$ (though not
monotonically), while the thermal pressure ($P_{\rm th}/k = 1.92 n_e
T$) rises from 20,000 to 59,000 K cm$^{-3}$.  This calculation assumes
that the \osix -emitting plasma has a constant emissivity per ion and
that scattering by non-emitting species is negligible.  If \nosix\ is
considered an upper limit, then $n_e$ and $P_{\rm th}/k$ are higher
than the derived values.

\section{DISCUSSION}

\citet{Shelton:01} detect \osix\ emission at 1032 and 1038 \AA\ with
intensities of $2930 \pm 290$ and $1790 \pm 260$ \lu, respectively, in
an $\sim$ 200 ksec \fuse\ observation centered on $l = 315.0, b =
-41.3$.  Though nearly antipodal, our sight lines show similar
\osix\ intensities, suggesting that diffuse \osix\ emission with an
observed doublet intensity of $\sim$ 5000 \lu\ may be a general feature
of high-Galactic-latitude sight lines.

Both the Coma and Virgo Clusters show up clearly in the \rosat\
\onequarter -keV surface-brightness map of \citet{Snowden:97}.  The
plasma responsible for the observed soft--X-ray (SXR) flux is likely to
be somewhat hotter than that producing our \osix\ emission.  Also
prominent in this map is the North Polar Spur (NPS), thought to be
gas shock-heated to $10^6$ K by stellar winds and supernovae from hot
stars in the Sco-Cen association \citep{Weaver:79, Heiles:84}.  The
Virgo Cluster lies near the edge of the SXR emission associated with
the NPS.

The infrared emission maps of \citet*{Schlegel:98} show that the Coma
Cluster lies in a region of the sky nearly free of dust, with a
reddening of only \ebv\ = 0.008.  Assuming the extinction
parameterization of \citet*{CCM:89} and $R_V = 3.1$, this reddening
corresponds to an attenuation of $\sim$ 10\% at 1035 \AA.  The infrared
sky is more complex in the direction of the Virgo Cluster, as the
cluster lies on the edge of a broad dust lane that may be associated
with the NPS.  The extinction toward Virgo is \ebv\ = 0.030.  Any
\osix\ $\lambda \lambda 1032, 1038$ emission originating beyond this
obscuring material will be reduced in intensity by $\sim$ 34\%
(hence our reddening correction of 1.5).

The column density \nosix\ derived from our M87 spectrum falls at the
low end of the range \nosix\ $\sim (1.4 - 5.0) \times 10^{14}$
cm$^{-2}$ obtained by \citet{Savage:00} for sight lines through the
Galactic halo.  Such a low value of \nosix\ suggests that our Virgo
sight line does not intercept significant \osix -emitting
plasma associated with the North Polar Spur.  If so, then parameters
derived for this sight line may be typical of the Galactic halo.  Given
that the total pressure (thermal + nonthermal) in the Galactic plane
may be as low as $\sim$ 25,000 K cm$^{-3}$ \citep{McKee:93},
the thermal pressure we derive for the halo seems quite high.  The
derived thermal pressure would rise if \nosix\ is lower than our
best-fit value, but would fall if the cirrus clouds responsible for
dust extinction lie beyond the \osix -emitting plasma.  Additional
\fuse\ observations of M87, already scheduled for Cycle 2, should
yield a greatly-improved determination of \nosix.

As seen in Table 1, the LSR velocities of the \osix\ 1032
and 1038 \AA\ emission features differ significantly from one another
in both the Coma and Virgo spectra.  Neither instrumental calibration
uncertainties nor contamination by emission from other species can
explain this effect.  More likely, the apparent velocity differences are
optical-depth effects.  Because the optical depth of the 1038 \AA\ line
is half that of the 1032 \AA\ line, the 1038 \AA\ flux that we observe
may have been emitted in regions far beyond those responsible for the
1032 \AA\ flux.  \osix\ component velocities from multiple sight
lines may thus be used to map velocity fields in the Galactic halo, 
even when individual velocity components are not resolved.

Our serendipitous detection of diffuse \osix\ emission bodes well both for
continued \fuse\ observations of selected sight lines and for proposed
missions, such as {\it SPEAR} ({\it Spectroscopy of Plasma Evolution from
Astrophysical Radiation;} \citealt*{Edelstein:00}), which seek to map
the entire sky at FUV wavelengths.

\acknowledgments

We thank R.\ Shelton and E.\ Murphy for discussions of \fuse\ data
analysis and acknowledge the outstanding efforts of the \fuse\
P.I.\ team to make this mission successful.  This research has made use
of NASA's Astrophysics Data System and is supported by NASA grant NAG
5-8956.



\end{document}